\documentclass[aps,prl,reprint,superscriptaddress,showpacs]{revtex4-1}

\usepackage{graphicx}
\usepackage{color}

\newcommand {\bfp} {{\bf p}}
\newcommand {\bfq} {{\bf q}}

\renewcommand {\d} {{\rm d}}

\newcommand {\E} {\varepsilon}

\newcommand {\om} {\omega}
\newcommand {\Om} {\Omega}

\begin{document}

\title{Revealing the Mechanism of Low-Energy Electron Yield Enhancement from Sensitizing Nanoparticles}

\author{Alexey V. Verkhovtsev}
\email[]{verkhovtsev@mail.ioffe.ru}
\affiliation{MBN Research Center, Altenh\"oferallee 3,
60438 Frankfurt am Main, Germany}
\affiliation{A.F. Ioffe Physical-Technical Institute,
Politekhnicheskaya ul. 26, 194021 St. Petersburg, Russia}

\author{Andrei V. Korol}
\email[]{korol@th.physik.uni-frankfurt.de}
\affiliation{MBN Research Center, Altenh\"oferallee 3,
60438 Frankfurt am Main, Germany}
\affiliation{Department of Physics, St. Petersburg State Maritime Technical University,
Leninskii prospekt 101, 198262 St. Petersburg, Russia}

\author{Andrey V. Solov'yov}
\email[]{solovyov@mbnresearch.com}
\affiliation{MBN Research Center, Altenh\"oferallee 3,
60438 Frankfurt am Main, Germany}
\affiliation{A.F. Ioffe Physical-Technical Institute,
Politekhnicheskaya ul. 26, 194021 St. Petersburg, Russia}


\begin{abstract}
We provide a physical explanation for enhancement of the low-energy electron
production by sensitizing nanoparticles due to irradiation by fast ions.
It is demonstrated that a significant increase in the number of emitted electrons
arises from the collective electron excitations in the nanoparticle.
We predict a new mechanism of the yield enhancement due to the plasmon excitations
and quantitatively estimate its contribution to the electron production.
Revealing the nanoscale mechanism of the electron yield enhancement, we provide an
efficient tool for evaluating the yield of emitted electron from various sensitizers.
It is shown that the number of low-energy electrons generated by the gold and
platinum nanoparticles of a given size exceeds that produced by the equivalent
volume of water and by other metallic (e.g., gadolinium) nanoparticles by an
order of magnitude.
This observation emphasizes the sensitization effect of the noble metal nanoparticles
and endorses their application in novel technologies of cancer therapy with ionizing
radiation.

\end{abstract}

\pacs{36.40.Gk, 79.20.-m, 61.80.-x, 87.53.-j} 

%

\maketitle


Radiotherapy is currently one of the frequently used methods to
treat cancer, which is a major health concern of nowadays.
However, it has the side effects induced by the radiation in surrounding
healthy tissues.
One of the promising modern treatment techniques is ion-beam cancer therapy 
\cite{Schardt_2010_RevModPhys.82.383, Durante_2010_NatRevClinOncol.7.37,
Surdutovich_2014_EPJD_Colloquia_Paper}.
It allows one to deliver a higher dose to the target region, as compared to
conventional photon therapy, and also to minimize the exposure of healthy
tissue to radiation \cite{Schardt_2010_RevModPhys.82.383}.
Approaches that enhance radiosensitivity within tumors relative to normal tissues
have the potential to become advantageous radiotherapies.
A search for such approaches is within the scope of several ongoing
multidisciplinary projects \cite{COST_Nano-IBCT, ARGENT}.

Metal nanoparticles (NPs) were proposed recently to act as sensitizers
in cancer treatments with ionizing radiation
\cite{Porcel_2010_Nanotechnology.21.085103,
Liu_2013_Nanoscale.5.11829, Porcel_2014_NanomedNanotechBiolMed}.
Injection of such nanoagents into a tumor can increase relative biological
effectiveness of ionizing radiation.
It is defined as a ratio of the dose delivered by photons to that
by a different radiation modality, leading to the same biological effects
such as the probability of an irradiated cell death.
During the past years, application of gold NPs acting as dose enhancers,
in combination with photons, revealed an increase of cancer cell killing
\cite{McMahon_2011_SciRep.1.18, Jain_2011_IntJRadiatOncolBiolPhys.79.531,
Hainfeld_2004_PhysMedBiol.49.N309},
while an advantage of using NPs in ion-beam cancer therapy is still to be
thoroughly substantiated.

It is currently acknowledged
\cite{Surdutovich_2014_EPJD_Colloquia_Paper, Michael_2000_Science.287.1603,
G.Garcia_RadDamage, Solov'yov_2009_PhysRevE.79.011909}
that a substantial portion of biodamage by incident ions is related
to the secondary electrons and free radicals produced due to ionization
of the medium by the projectiles.
Refs. \cite{Boudaiffa_2000_Science.287.1658, Huels_2003_JAmChemSoc.125.4467,
Toulemonde_2009_PhysRevE.80.031913}
have explored the possibility of the low-energy electrons (LEE), having the
kinetic energy from a few eV to several tens of eV, to be important agents
of biodamage.

In this Letter, we reveal the physical mechanism of enhancement of the LEE
production by sensitizing (noble metal, in particular) NPs.
We demonstrate that a significant increase in the number of emitted electrons
due to irradiation by fast ions comes from the two distinct types of collective
electron excitations.
We predict that the yield of the $1-10$~eV electrons is strongly enhanced due to
the decay of plasmon-type excitations of delocalized valence electrons in metal NPs.
More specifically, the leading mechanism of the electron production is associated
with the {\it surface} plasmon, whose contribution to the electron yield exceeds
by an order of magnitude that of the volume plasmon, considered in the recent
Monte Carlo simulation \cite{Waelzlein_2014_PhysMedBiol.59.1441}.
For higher electron energies (of about $10-30$~eV), the dominating
contribution to the electron yield arises from the atomic giant resonances
associated with the collective excitation of $d$ electrons in individual
atoms in a NP.
As a result of these effects, the number of the LEE generated by the noble metal NP
of a given size exceeds that produced by the equivalent volume of water by an order
of magnitude.
Based on the physical understanding of the processes involved, we provide an efficient
tool for a quantitative estimate of the yield of emitted electrons from sensitizing NPs.


Studying the electron production by a NP irradiated by ions,
we account for the two collective electron effects, namely
excitation of delocalized electrons in a NP (plasmons) and
that of $d$ electrons in individual atoms (atomic giant resonances).
These phenomena occur in various processes of interaction of ionizing
radiation with matter.
In particular, dipole collective excitations result in the formation of prominent
resonances in the photoabsorption spectra of atomic clusters and nanoparticles
\cite{Kreibig_Vollmer, Suraud_2013_ClusterScience},
while the impact ionization cross sections comprise also the contributions
of higher multipole terms \cite{Gerchikov_1997_JPhysB.30.5939}.

We consider noble metal (gold, platinum, and silver) and other metallic
(gadolinium) NPs, which are of interest
\cite{Porcel_2010_Nanotechnology.21.085103,
Liu_2013_Nanoscale.5.11829, Porcel_2014_NanomedNanotechBiolMed,
McMahon_2011_SciRep.1.18, Jain_2011_IntJRadiatOncolBiolPhys.79.531}
in application in cancer treatments.
As a starting point, we have calculated the photoabsorption spectra of
several 3D gold clusters made of 18 to 42 atoms.
The calculations within the time-dependent density-functional theory
(TDDFT) framework
\cite{Runge_Gross_1984_PhysRevLett.52.997, Walker_2006_PhysRevLett.96.113001}
were performed using the Quantum Espresso package
\cite{Giannozzi_2009_JPhysCondMat.21.395502, Malcioglu_2011_CompPhysCommun.182.1744}.
%
A technical description of the calculations is presented in the Supplementary Material.
%
As a case study, Fig.~\ref{fig_Au32_plasmon} presents the TDDFT-based
spectrum of the Au$_{32}$ cluster in the photon energy range up to
60~eV (thin black curve).
The spectrum is characterized by a low-energy peak located below 10~eV
and by a broad feature with a maximum at about 25~eV.
The results of the calculation are compared to the X-ray absorption data
for atomic gold \cite{Henke_1993_AtDataNuclDataTables.54.181},
multiplied by the number of atoms in the cluster.

\begin{figure}[ht]
\centering
\includegraphics[width=0.43\textwidth,clip]{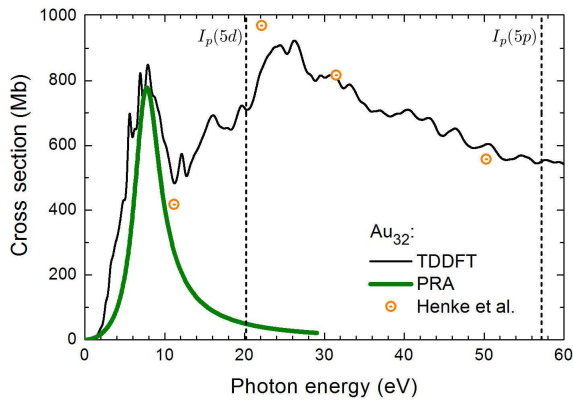}
\caption{(color online).
Photoabsorption cross section of the Au$_{32}$ cluster calculated within
TDDFT (thin curve).
Thick curve represents the contribution of the plasmon-type excitations.
Symbols represent the data for atomic gold
\cite{Henke_1993_AtDataNuclDataTables.54.181}, multiplied by the number
of atoms in the cluster.
Vertical lines mark the $5d$ and $5p$ ionization thresholds in the atom
of gold.}
\label{fig_Au32_plasmon}
\end{figure}

Our analysis has revealed that the high-energy feature is the atomic giant resonance
formed due to the excitation of electrons in the $5d$ atomic shell.
The integration of the oscillator strength from 20.2~eV
(ionization threshold of the $5d$ shell in the atom of gold) up to 57.2~eV
(the $5p$ shell ionization threshold \cite{Henke_1993_AtDataNuclDataTables.54.181}),
indicates that about eight localized $d$-electrons contribute to the excitation
of the $5d$ shell forming the broad peak in the spectrum.
The low-energy peak is due to the plasmon-type excitation, which involves some
fraction of $s$ and $d$ electrons delocalized over the whole cluster.
The delocalization comes from a partial hybridization of the $6s$ and $5d$
atomic shells.
The integration of the oscillator strength up to 11.2~eV (energy at which the
first dip after the resonance peak is observed in the TDDFT spectrum) reveals
that about 1.5 electrons from each atom contribute to the collective plasmon-type
excitation,
%
indicating delocalization of some fraction of $d$ electrons
\cite{Fernandez_2006_PhysRevB.73.235433}.
Thus, the total photoabsorption spectrum of a gold NP in the energy region up
to 60~eV approximately is equal to the sum of the plasmon contribution and that
of the $5d$ electron excitations in individual atoms,
$\sigma_{\gamma} \approx \sigma_{\rm pl} + \sigma_{\rm 5d}$.

Similar to the photoionization, the two distinct types of collective
electron excitations appear in the process of impact ionization.
We use the methodology allowing us to analyze the role of these contributions
to the electron production by sensitizing NPs separately.
The single differential inelastic scattering cross section of a fast projectile
in collision with a NP is given by a general formula
(we use the atomic system of units, $m_e = |e| = \hbar = 1$):
\begin{equation}
\frac{{\rm d}\sigma}{{\rm d}\Delta\E} =
\frac{2\pi}{p_1 p_2} \int\limits_{q_{\rm min}}^{q_{\rm max}} q \, {\rm d}q
\frac{{\rm d}^2\sigma}{{\rm d}\Delta \varepsilon \, {\rm d}\Omega_{{\bf p}_2}}
\approx
\frac{{\rm d}\sigma_{\rm pl}}{{\rm d}\Delta\E} +
\frac{{\rm d}\sigma_{\rm at}}{{\rm d}\Delta\E}  \ ,
\label{eq_01}
\end{equation}

\noindent
where
$\Delta \varepsilon = \E_1 - \E_2$ is the energy loss of the incident projectile
of energy $\E_1$,
${\bfp}_1$ and ${\bfp}_2$ the initial and the final momenta of the projectile,
$\Om_{{\bfp}_2}$ its solid angle,
and ${\bfq} = {\bfp}_1 - {\bfp}_2$ the transferred momentum.
The cross sections ${\rm d}\sigma_{\rm pl}$ and ${\rm d}\sigma_{\rm at}$ denote
the contributions of the plasmon and individual atomic excitations, respectively.

The contribution of the plasmon excitations to the ionization cross section is
described by means of the plasmon resonance approximation (PRA)
\cite{Kreibig_Vollmer, Connerade_AS_PhysRevA.66.013207,
Solovyov_review_2005_IntJModPhys.19.4143, Verkhovtsev_2012_EPJD.66.253},
which postulates that
the collective excitations dominate the cross section in the vicinity of
the plasmon resonance.
During the past years, the PRA was successfully applied to explain the
resonant-like structures in photoionization spectra
\cite{Connerade_AS_PhysRevA.66.013207, Verkhovtsev_2013_PhysRevA.88.043201}
and differential inelastic scattering cross sections
\cite{Solovyov_review_2005_IntJModPhys.19.4143, Mikoushkin_1998_PhysRevLett.81.2707,
Verkhovtsev_2012_JPhysB.45.141002} 
of metal clusters and carbon fullerenes by the photon and electron impact.
Within the PRA,
the double differential cross section $\d^2\sigma / \d\Delta\E \, \d\Om_{{\bfp}_2}$
for a spherical NP is defined as a sum of the surface ($s$) and the volume
($v$) plasmon terms, which are constructed as a sum over different multipole
contributions corresponding to different values of the angular momentum $l$
\cite{Solovyov_review_2005_IntJModPhys.19.4143}:
\begin{eqnarray}
\begin{array}{l l}
\displaystyle{ \frac{\d^2\sigma^{(s)} }{\d\Delta\E \, \d\Om_{{\bfp}_2}} }
\propto
\sum\limits_{l}
\frac{ \om_{l}^{(s)2}\, \Gamma_{l}^{(s)}  }
{ \bigl(\om^2-\om_{l}^{(s)2}\bigr)^2 + \om^2\Gamma_{l}^{(s)2} }
\vspace{0.2cm} \\
\displaystyle{ \frac{\d^2\sigma^{(v)} }{\d\Delta\E \, \d\Om_{{\bfp}_2}} }
\propto
\sum\limits_{l}
\frac{ \om_p^2\, \Gamma_l^{(v)} }
{ \bigl(\om^2-\om_p^2\bigr)^2+\om^2\Gamma_l^{(v)2} } \ .
\end{array}
\label{Equation.02}
\end{eqnarray}

\noindent
Here $\om_{l}^{(s)} = \sqrt{ l / (2l+1) } \, {\om_p}$ is the frequency of the
surface plasmon of the multipolarity $l$, and
$\om_p = \sqrt{4 \pi \rho_0} = \sqrt{ 3N_e / R^3}$ is the volume plasmon
frequency associated with the density $\rho_0$ of $N_e$ delocalized electrons.
The quantities $\Gamma_l^{(i)}$ ($i = s,v$) are the plasmon widths.
%
In the analysis, we accounted for the dipole ($l= 1$), quadrupole ($l = 2$)
and octupole ($l = 3$) terms.
Excitations with larger $l$ have a single-particle rather than a collective
nature \cite{Solovyov_review_2005_IntJModPhys.19.4143}, thus not contributing
to the plasmon formation.
Explicit expressions for the cross sections (\ref{Equation.02}), obtained
within the first Born approximation, are presented in Ref. \cite{Verkhovtsev_2012_EPJD.66.253}.
This approach is applicable for the collision of a nanoparticle with a
fast heavy projectile.

The PRA relies on a few parameters, which include the oscillator strength
of the plasmon excitation, position of the peak and its width.
In the dipole case, these were validated by fitting the TDDFT-based spectra
of several 3D gold clusters to those calculated within the model approach.
We assumed that 1.5 electrons from each gold atom contribute to the plasmon excitation.
This value, along with the dipole plasmon width
$\Gamma_1^{(s)} = 4.0$~eV $\approx 0.6 \, \om_1^{(s)}$,
was used to reproduce the low-energy peak in the photoabsorption spectra of
gold clusters by means of the PRA scheme
(see the solid green curve in Fig.~\ref{fig_Au32_plasmon}).
The similar ratio of the plasmon resonance width to its frequency was assumed
for higher multipole terms of the surface plasmon \cite{Note_plasmon_width} and
for the volume plasmon as well,
$\Gamma_l^{(s)}/\om_l^{(s)} = \Gamma_l^{(v)}/\om_p = 0.6$.


Atomic $d$ electrons in noble metals play a dominant role at higher excitation
energies, from approximately 20 to 60~eV (see Fig.~\ref{fig_Au32_plasmon}
for the case of gold).
For distant collisions,
i.e. when the impact parameter exceeds the radius $R$ of the atomic subshell,
the ionization spectra are dominated by the dipole term
\cite{Landau_Lifshitz_3}.
Comparing the cross sections of photoionization, $\sigma_{\gamma}$, and
the dipole term of inelastic scattering, ${\rm d}\sigma_{\rm at} / {\rm d}\Delta\E$,
calculated in the Born approximation,
one derives the following expression:
\begin{equation}
\frac{{\rm d}\sigma_{\rm at}}{{\rm d}\Delta\E} = \frac{2c}{\pi \om v_1^2} \sigma_{\gamma}
\ln{ \left( \frac{v_1}{\om R} \right)} \ ,
\label{ElScatter_to_PI}
\end{equation}

\noindent
where 
$\om = \E_1 - \E_2$ is the energy transfer, and
$v_1$ the projectile velocity.
%
Equation~(\ref{ElScatter_to_PI}), obtained within the so-called
''logarithmic approximation'', assumes that the main contribution to
the cross section ${\rm d}\sigma_{\rm at} / {\rm d}\Delta\E$ comes from
the region of large distances, $R < r < v_1/\om$.
This relation has the logarithmic accuracy which implies that the
logarithmic term dominates the cross section while all non-logarithmic
terms are neglected \cite{Korol_AVS_BrS_2014}.
%
For the studied noble metal atoms, we assumed $\om \approx 1$~a.u.
which corresponds to the maximum of the $5d \, (4d)$ giant resonance in
gold and platinum (silver) \cite{Henke_1993_AtDataNuclDataTables.54.181},
$v_1 \approx 6.3$~a.u. for a 1~MeV proton,
and the electron shell radii
$R_{5d}({\rm Au, Pt}) \approx R_{4d}({\rm Ag}) \approx$~2~a.u.
Note that the interaction of the incident projectile with the NP leads to
the formation of the giant resonance not in all atoms of the system but
only in those located within the impact parameter interval from
$r_{\rm min} \simeq R_{5d} \, (R_{4d})$ to $r_{\rm max} \simeq v_1/\om$.
This estimate reveals that the resonance is excited in approximately one
third of the atoms.
A similar estimate was also made for a Gd atom.
Contrary to the noble metals, the Gd atom has a single electron in the $5d$ shell.
Thus, there is no atomic giant resonance in the ionization spectrum of Gd in
the $20-60$~eV range, and
the spectrum is characterized by a narrow peak at $\om \approx 1.2$~a.u.,
formed due to ionization of the $5p$ shell.

To quantify the production of secondary electrons in collision with the
nanoparticles, the cross section ${\rm d}\sigma / {\rm d}\Delta\E$,
Eq.~(\ref{eq_01}), is redefined as a function of the kinetic energy $E$
of the electrons:
$E = \Delta\E - I_p$, where
$I_p$ is the ionization threshold of the system.
The cross section ${\rm d}\sigma / {\rm d}E$ can be related
to the probability to produce $N$ electrons with kinetic energy within the
interval $dE$, emitted from a segment ${\rm d}x$ of the trajectory,
via \cite{Surdutovich_2014_EPJD_Colloquia_Paper}:
\begin{equation}
\frac{{\rm d}^2 N(E)}{{\rm d}x \,{\rm d}E} = n \frac{{\rm d}\sigma}{{\rm d}E} \ ,
\end{equation}

\noindent
where $n$ 
is the atomic density of the target.

\begin{figure}[ht]
\centering
\includegraphics[width=0.43\textwidth,clip]{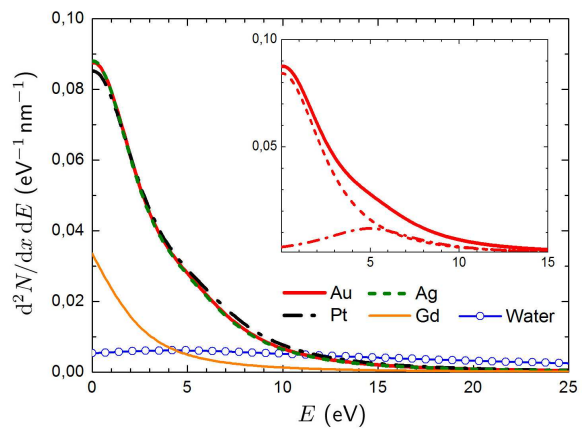}
\caption{(color online).
Number of electrons per unit length per unit energy produced via the plasmon
excitations in the Au, Pt, Ag and Gd NPs irradiated by a 1~MeV proton.
Open circles represent the number of electron generated from the equivalent
volume of water \cite{deVera_2013_PhysRevLett.110.148104}.
Inset: contributions of the surface (dashed) and the volume (dash-dotted)
plasmons to the electron yield from the AuNP.}
\label{fig_Au_ElProd_2}
\end{figure}

Figure~\ref{fig_Au_ElProd_2} presents the number of electrons per
unit length per unit energy produced via the {\it plasmon excitation}
mechanism by the 1~nm spherical NPs due to 1~MeV proton irradiation.
We have also compared the electron production by the NPs and by the
equivalent volume of pure water medium \cite{deVera_2013_PhysRevLett.110.148104}.
Comparative analysis of the spectra demonstrates that the number of LEE
(with the kinetic energy of about a few eV) produced due to the plasmon
excitations in the noble metal NPs is about one order of magnitude higher
than that by liquid water.
The enhancement of the LEE yield due to the presence of sensitizing NPs
may increase the probability of the tumor cell destruction due to the higher
number of double and multiple strand breaks of the DNA.
Thus, the plasmon decay in the nanoparticles, embedded in a biological medium,
represents an important channel for production of low-energy secondary electrons
in the medium.

It was demonstrated recently \cite{Porcel_2014_NanomedNanotechBiolMed}
that tumor-targeted Gd-based NPs amplify cell death under ion irradiation
and also enhance the number of single and double strand breaks in plasmid DNA.
However, it was noted that the effect of GdNPs is less pronounced than
that of platinum-based compounds. 
This result generally corresponds to out conclusions that the electron
yield from a GdNP exceeds the electron production from pure water medium
but is lower than that from noble metal NPs.

The low electron yield from the GdNP, as compared to the noble metal
targets, is explained by the density effects (the atomic density of Gd is about
two times smaller than that of the studied noble metals)
as well as by the lower plasmon frequency.
The maximum of the plasmon resonance peak in the GdNP (4.1~eV) is
located below the ionization potential of the system ($\sim 5.0$~eV)
\cite{Yuan_2014_JChemPhys.140.154308}.
In the case of noble metal NPs, the plasmon peak maxima are in the range between
5.5 and 6.0~eV, being in the vicinity of the ionization thresholds.
Therefore, the plasmon decay in noble metal NPs results in the more intense electron
emission as compared to the GdNP.
In the latter case, the plasmon will mostly decay into the single-electron excitations,
which can lead to the vibration of the ionic core as a result of the electron-phonon
coupling \cite{Gerchikov_2000_JPhysB.33.4905}.

The inset of Fig.~\ref{fig_Au_ElProd_2} demonstrates that the {\it surface} plasmon
(dashed curve) gives the dominating contribution to the electron production by the
metallic NP (as a case study, we consider gold) exceeding that of the volume plasmon
(dash-dotted curve) by an order of magnitude.
The significance of the plasmon excitations in the process of electron production by
sensitizing NPs, revealed in this work, repudiates the statement made in
Ref.~\cite{Waelzlein_2014_PhysMedBiol.59.1441} on the negligible role of the plasmon
excitations in forming the spectrum of emitted electrons.
Let us stress that only the volume plasmon excitation was accounted for in
the cited paper.
%
Our more extended analysis \cite{Gold_el_prod_extended} reveals that
the plasmon excitations play a prominent role in the production
of LEE from gold NPs of about $1 - 5$ nm in diameter.
This size range corresponds to the size of noble metal and Gd-based NPs studied
recently in relation to the radiotherapies with charged ions
\cite{Porcel_2010_Nanotechnology.21.085103, Porcel_2014_NanomedNanotechBiolMed}.

\begin{figure}[ht]
\centering
\includegraphics[width=0.43\textwidth,clip]{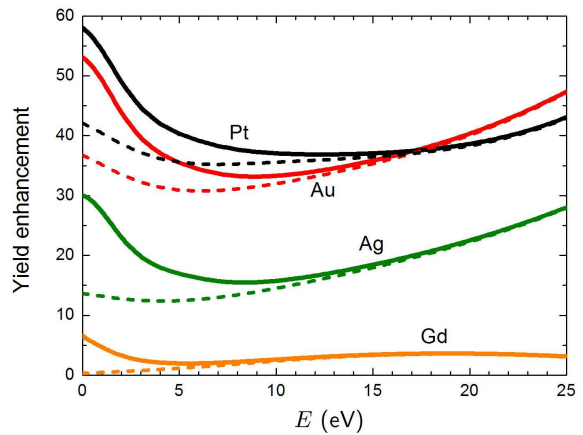}
\caption{(color online).
Yield enhancement from the 1~nm metallic NPs.
Dashed lines show the contribution of individual atomic excitations.
Solid lines show the resulting contribution with an account of the plasmons.}
\label{fig_Au_ElProd_3}
\end{figure}

To estimate the total number of electrons produced due to the collective excitations
in the NPs, we have also accounted for the contribution of excitations in individual
atoms.
Figure~\ref{fig_Au_ElProd_3} demonstrates the relative enhancement of the electron
yield from the considered nanoparticles as compared to pure water.
This quantity was obtained by summing up the contribution of the plasmons and
individual atomic excitations.
The dashed lines present the contribution of the atomic giant resonances
($5d$ in Au and Pt, and $4d$ in Ag) as well as the total $5p+5d$ contribution
in Gd, estimated using Eq.~(\ref{ElScatter_to_PI}).
Making this estimate, we have assumed that the ionization cross sections are
dominated by the dipole excitation.
Contribution of quadrupole and higher multipole terms will lead to an
increase in the number of emitted electrons but their relative contribution
will be not as large as that from the dipole excitation.
The solid line is the sum of the excitations in individual atoms and the plasmons.
The significant yield enhancement arises in those nanoparticles whose constituent
atoms possess the giant resonance, contrary to case of gadolinium which has a
single $5d$ electron.
Accounting for the plasmon contribution leads to a significant increase of the
$1-5$~eV electron yield.
Due to the collective electron excitations arising in these systems, the gold
and platinum NPs can thus produce much larger (of about an order of magnitude)
number of LEE comparing to the equivalent volume of pure water medium.
%
We note that the enhanced production of LEE will also lead to an increase
in the number of free radicals as well as other reactive species, like hydrogen peroxide
${\rm H}_2{\rm O}_2$, which can travel the distances larger than the cell
nucleus \cite{Porcel_2014_NanomedNanotechBiolMed}.
Thus, these species can deliver damaging impacts onto the DNA from the radiation
induced damages associated with the presence of NPs in other cell compartments,
such as lysosomes \cite{Stefancikova_2014_CancerNanotech.5.6}.


To conclude, we have 
analyzed the electron production by sensitizing metallic nanoparticles
due to irradiation by fast ions and revealed the physical mechanism of
low-energy electron yield enhancement.
It has been shown that the significant increase in the number of emitted
electrons arises from the two distinct types of collective electron excitations
formed in nanoparticles.
The yield enhancement is caused by the plasmons, excited in a whole nanoparticle,
and by the excitation of $d$ electrons in individual atoms that results in the
formation of giant atomic resonances.
%
%
The plasmon excitation mechanism leads a significant additional increase of the
electron yield enhancement from the noble metal nanoparticles comparing to water.
Thus, the damping of the plasmons excited in the metal nanoparticles represents
an important mechanism of the low-energy electron generation.


In this Letter, we have introduced a general methodology which can be
applied for other nanoscale systems proposed as sensitizers in cancer therapy.
Particularly, it can be applied to study more complex types of sensitizers,
for instance core-shell nanoparticles, where the collective electron
excitations will arise in both parts of the system.
A proper choice of the constituents will allow one to tune the position of
the resonance peaks in the ionization spectra of such systems and, subsequently,
to cover a broader kinetic energy spectrum of electrons emitted from these
nanoparticles.
%
The utilized methodology can also be adopted for different projectiles,
e.g. carbon ions, which are the most clinically used projectiles,
besides protons.
%


We are grateful to Pablo de Vera for providing us with the data on electron
production in pure water and acknowledge the Frankfurt Center for Scientific
Computing (CSC) for providing computer facilities.


\end{document}